\documentclass[preprint]{revtex4}
\usepackage{graphicx}
\usepackage{dcolumn}
\usepackage{bm}

\begin{document}
\bibliographystyle{prsty}


\title{Anisotropy-induced Fano resonance}
\author{Cheng-Wei Qiu$^{1*}$, Andrey Novitsky$^{2}$, Lei Gao $^3$, Jian-Wen Dong$^4$ and Boris Luk'yanchuk$^{1,5\dagger}$}
\affiliation{$^{1}$Department of Electrical and Computer
Engineering, National University of Singapore, 4 Engineering Drive
3, Singapore 117576} \email{eleqc@nus.edu.sg}
 \affiliation{$^{2}$Department of
Photonics Engineering, Technical University of Denmark, Lyngby 2800,
Denmark} \affiliation{$^{3}$Jiangsu Key Laboratory of Thin Films,
Department of Physics, Soochow University, Suzhou 215006, China}
\affiliation{$^{4}$State Key Laboratory of Optoelectronic Materials
and Technologies, Sun Yat-Sen (Zhongshan) University, Guangzhou
510275, China} \affiliation{$^{5}$Data Storage Institute, DSI
Building, 5 Engineering Drive 1, Singapore,
117608}\email{Boris_L@dsi.a-star.edu.sg}


\date{\today}

\begin{abstract}
An optical Fano resonance, which is caused by birefringence control
rather than frequency selection, is discovered. Such
birefringence-induced Fano resonance comes with fast-switching
radiation. The resonance condition $\varepsilon _t< 1/\varepsilon _r
$ is revealed and a tiny perturbation in birefringence is found to
result in a giant switch in the principal light pole induced near
surface plasmon resonance. The loss and size effects upon the Fano
resonance have been studied Fano resonance is still pronounced, even
if the loss and size of the object increase. The evolutions of the
radiation patterns and energy singularities illustrate clearly the
sensitive dependence of Fano resonance upon the birefringence.

\end{abstract}

\maketitle

\newpage

Light scattering by a small particle is one of the most fundamental
problems in electrodynamics and potential applications in
information processing, nanotechnologies and engineering. The
essence of extraordinary light scattering is the localized surface
plasmon which is oscillating with the frequency of the incident
wave~\cite{Tsai}. For particles with the size much smaller than the
incident wavelength, Rayleigh approximation can be
adopted~\cite{Hulst}. Nonetheless, recent studies show that for
small particle with weak dissipation near plasmon resonance
frequencies, anomalous light scattering takes place with unexpected
features, e.g., sharp giant optical resonances with inverse
hierarchy, unusual frequency and particle size dependence, and
complicated near-field energy
circulation~\cite{Tribelsky,Luk'yanchuk1,Qiu}. Recently, a lot of
studies were related to the resonances in plasmonic
nanostructures~\cite{Maier_ACS09,Giessen09,Tribelsky2,Boris_NM},
coherent nanocavities~\cite{Maier_NL}, and metallic
films~\cite{Stockman,Garcia}. However, they were discussed for
isotropic materials or elements, and also Fano resonance was
observed versus the fine tuning of frequency.

In this letter, we provide a new paradigm to realize Fano resonance
via the birefringence (anisotropy) of a single particle, instead of
controlling frequency. This novel resonant mechanism may be a new
paradigm for sensitive optical identification of molecular groups,
calculation of heating, radiation pressure and trapping. The
anisotropy is found to be able to tailor the surface plasmon
resonance and induce additional plasmonic resonances. It is thus
revealed that birefringence-induced Fano resonance occurs to the
anisotropic rod and its radiation pattern is affected by the subtle
perturbation of the rod's birefringence. We also look into the near
field where Poynting bifurcation and vortex analysis are
investigated against the birefringence-induced Fano resonant cases.
Note that the anisotropic parameters are homogeneous, i.e.,
position-independent, which is in contrast to the parameters of a
cloak~\cite{Pendry}. Here we would like to consider the homogeneous
rod with constant radial anisotropy in both
$\mathord{\buildrel{\lower3pt\hbox{$\scriptscriptstyle\leftrightarrow$}}
\over \varepsilon } $
  and $\mathord{\buildrel{\lower3pt\hbox{$\scriptscriptstyle\leftrightarrow$}}
\over \mu } $, i.e., they are diagonal in cylindrical coordinates
with values $\varepsilon _r (\mu _r )$ in the radial
($\mathord{\buildrel{\lower3pt\hbox{$\scriptscriptstyle\rightharpoonup$}}
\over r}$) direction and $\varepsilon _t  (\mu _t )$ in the other
two directions
($\mathord{\buildrel{\lower3pt\hbox{$\scriptscriptstyle\rightharpoonup$}}
\over \theta} $ and
$\mathord{\buildrel{\lower3pt\hbox{$\scriptscriptstyle\rightharpoonup$}}
\over z}$). Actually, such birefringence can be realized by
graphitic multishells~\cite{Lucas} or stratified
mediums~\cite{Sten}, and in practice, it has been found in
phospholipid vesicle systems~\cite{Lange,Peterlin} and in cell
membranes containing mobile charges~\cite{Sukhorukov,Ambjornsson}.

 The magnetic field only exists in the
\emph{z}-direction. The constitutive tensors of the relative
permittivity and permeability are expressed as
$\mathord{\buildrel{\lower3pt\hbox{$\scriptscriptstyle\leftrightarrow$}}
\over \varepsilon }  = \varepsilon _r
\mathord{\buildrel{\lower3pt\hbox{$\scriptscriptstyle\rightharpoonup$}}
\over r}
\mathord{\buildrel{\lower3pt\hbox{$\scriptscriptstyle\rightharpoonup$}}
\over r}  + \varepsilon _t
\mathord{\buildrel{\lower3pt\hbox{$\scriptscriptstyle\rightharpoonup$}}
\over \theta }
\mathord{\buildrel{\lower3pt\hbox{$\scriptscriptstyle\rightharpoonup$}}
\over \theta }  + \varepsilon _t
\mathord{\buildrel{\lower3pt\hbox{$\scriptscriptstyle\rightharpoonup$}}
\over z}
\mathord{\buildrel{\lower3pt\hbox{$\scriptscriptstyle\rightharpoonup$}}
\over z} $ and
$\mathord{\buildrel{\lower3pt\hbox{$\scriptscriptstyle\leftrightarrow$}}
\over \mu }  = \mu _r
\mathord{\buildrel{\lower3pt\hbox{$\scriptscriptstyle\rightharpoonup$}}
\over r}
\mathord{\buildrel{\lower3pt\hbox{$\scriptscriptstyle\rightharpoonup$}}
\over r}  + \mu _t
\mathord{\buildrel{\lower3pt\hbox{$\scriptscriptstyle\rightharpoonup$}}
\over \theta }
\mathord{\buildrel{\lower3pt\hbox{$\scriptscriptstyle\rightharpoonup$}}
\over \theta }  + \mu _t
\mathord{\buildrel{\lower3pt\hbox{$\scriptscriptstyle\rightharpoonup$}}
\over z}
\mathord{\buildrel{\lower3pt\hbox{$\scriptscriptstyle\rightharpoonup$}}
\over z} $ respectively in cylindrical coordinates,  where
$\varepsilon _r (\mu _r )$
 and $\varepsilon _t  (\mu _t  )$
  stand for the permittivity (permeability) elements
corresponding to the electric- and magnetic- field vector normal to
and tangential to the local optical axis, respectively. The time
dependence $e^{ - i\omega t} $ is assumed. We rely on
Mathematica$^{TM}$ to produce all the results throughout.

For the TE mode, we have the wave equation
\begin{eqnarray}
\frac{1}{r}\left[ {\frac{\partial }{{\partial r}}\left(
{\frac{r}{{\varepsilon _t  }}\frac{{\partial H_z }}{{\partial r}}}
\right)} \right] + \frac{1}{{r^2 }}\frac{\partial }{{\partial
\theta }}\left( {\frac{1}{{\varepsilon _r }}\frac{{\partial H_z
}}{{\partial \theta }}} \right) + k_0^2 \mu _t  H_z  = 0.
\end{eqnarray}
The local field solutions in the inner and outer region of the wire
can be described as $ H_z^{in} = \sum\limits_{m =  - \infty }^\infty
{i^m A_m^{} J_{m'} (kr)e^{im\theta } }$ and $ H_z^{out} =
\sum\limits_{m = - \infty }^\infty  {i^m [J_m (k_0 r) + B_m^{}
H_m^{(1)} (k_0 r)]e^{im\theta } } $, respectively, where $k =k_0
\sqrt {\varepsilon _t  \mu _t  } $  and $k_0  = \omega \sqrt
{\varepsilon _0 \mu _0 } $. Note that the Bessel function index
\emph{m'}, i.e., $m'^2 = m^2 \varepsilon_t/\varepsilon_r $ is no
longer a conventional integer, which is different from the uniaxial
material considered in \cite{Bohren}. Under TE (or TM) incidence,
the uniaxial material in \cite{Bohren} is actually isotropic, while
our birefringent material is defined in cylindrical coordinate and
is not isotropic under TE (or TM) incidence
\cite{Lucas,Sten,Lange,Peterlin,Sukhorukov,Ambjornsson}. To solve
the scattering problem, the scattering coefficient $B_m $ now
becomes the most important issue:
\begin{equation}
B_m  =  - \frac{\sqrt{\varepsilon_t\mu_t} J_{m'}(\sqrt {\varepsilon
_t \mu_t}q)J_{m}^{'}(q)-\mu_t
J_m(q)J_{m'}^{'}(\sqrt{\varepsilon_t\mu_t}q)}{ {\sqrt {\varepsilon
_t  \mu _t} J_{m'}(\sqrt {\varepsilon_t
\mu_t}q)H_{m}^{(1)'}(q)-\mu_t
J_{m'}^{'}(\sqrt{\varepsilon_t\mu_t}q)}H_m^{(1)}(q)}
\end{equation}
where the prime denotes the derivative with respect to the argument
and the size parameter is $\emph{q}=k_0a$. It is evident that $B_m $
reduces to the isotropic case by replacing $\varepsilon_r
(\varepsilon_t)$ and $\mu_r (\mu_t)$ with $\varepsilon$ and $\mu$,
respectively~\cite{Bohren}.

Note that the ``effective'' permittivity and permeability can be
isotropic even for a radially birefringent object~\cite{Fung}. In
search of the effective response for such a birefringent cylinder,
one assumes that the rod with radial anisotropy is embedded in an
effective medium with isotropic effective permittivity $\varepsilon
_{eff} $ and permeability $\mu _{eff}$ and when no scattering
arises, the search stops, i.e., ``effective'' parameters are found.
In this sense, we replace $\varepsilon _0 $ and $\mu _0 $ by
$\varepsilon _0\varepsilon _{eff} $ and $\mu _0\mu _{eff} $, and
they can be determined. By considering the equivalence to the
isotropic case for $q \to 0$, the surface resonant condition tends
to be $\varepsilon _{eff}  = \sqrt {\varepsilon _r}
\sqrt{\varepsilon _t } = - 1$, i.e., $\varepsilon _t= 1/\varepsilon
_r $ for negative $\varepsilon_r$.

To explain the optical resonances numerically, let us present the
amplitude
\begin{equation}
B_m  =  - \frac{{\Re _m }}{{\Re _m  + i\Im _m }} \end{equation} by
separating the real and imaginary parts, then we have
\begin{equation}
\Re _m = \sqrt {\varepsilon _t  \mu _t  } J_{m'} (\sqrt {\varepsilon
_t \mu _t } q)J_m '(q) - \mu _t  J_m (q)J_{m'} '(\sqrt {\varepsilon
_t \mu _t } q) \end{equation} \begin{equation}\Im _m  = \sqrt
{\varepsilon _t \mu _t } J_{m'} (\sqrt {\varepsilon _t  \mu _t  }
q)Y_m '(q) - \mu _t J_{m'} '(\sqrt {\varepsilon _t \mu _t  } q)Y_m
(q) \end{equation} where $Y_m $ is the Neumann function. The exact
optical resonance corresponds to the situation when $\Im _m  = 0$,
which leads to $\left| {B_m } \right| = 1$. For simplicity, the
material is assumed to be non-magnetic, i.e., $\mu _t = 1$.

\begin{figure}
\centerline{\includegraphics[width=9cm]{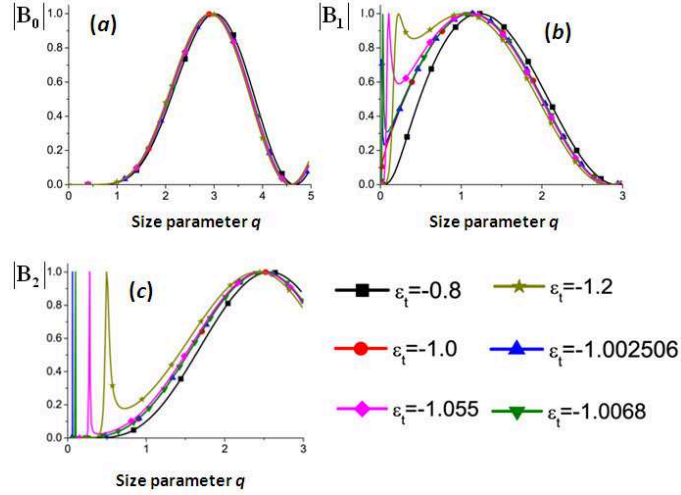}}
\vspace{-2ex}\caption{(Color online) Analysis of resonant conditions
and dominant modes versus size parameter \emph{q} for monopole
resonance \emph{m}=0 (a); dipole resonance \emph{m}=1 (b), and
quadrupole resonance \emph{m}=2 (c). (a-c) show how the new
additional resonances will be induced by the tiny perturbation in
birefringence for high-order resonances (e.g., dipole, quadrupole,
etc). Here we keep $\varepsilon_r = - 1$, and
$\delta=\varepsilon_t-\varepsilon_r$ which serves the measure of
birefringence. Hence $\varepsilon _t = - 1$ (i.e., $\delta=0$)
corresponds to the isotropic situation. } \label{fig:1}
\end{figure}
In Fig.~\ref{fig:1}(a), it indicates that there is no surface
resonance at small \emph{q} and the maximum magnitude of $B_0$
always occurs in the vicinity of $q\approx 2.8$ no matter how the
transversal permittivity $\varepsilon_t$ deviates from the radial
permittivity $\varepsilon_r$. Nevertheless, new resonances will be
excited in higher-order modes (e.g., dipolar, quadrupolar, etc) as
shown in Figs.~\ref{fig:1}(b,c) at small size parameter \emph{q}. In
Fig.~\ref{fig:1}(c), one can see the isotropic case (the red line),
i.e., $\varepsilon_t=-1=\varepsilon_r$, does not have the additional
optical surface resonance at very small \emph{q}. Under the resonant
condition $\varepsilon_t<\varepsilon_r$, Figs.~\ref{fig:1}(b,c)
reveal that a slight deviation from the isotropic case
($\varepsilon_t=-1$), i.e., the perturbation of birefringence
$\delta\neq 0$, will give rise to a new surface resonance for small
particles. When $\varepsilon_t$ is very closer to $-1$, the
additional optical resonance for higher-order modes will be more
isolated from the volume resonance (which is still found to be
insensitive as the monopole). When $\varepsilon_t$ is more deviated
from $-1$, the optical resonance will be merging with the volume
resonance. Obviously, the case of $\varepsilon_t=0.8$ in
Fig.~\ref{fig:1} does not satisfy the surface resonant condition, so
it only possesses a volume resonance. Considering the resonance
trajectory, it is not difficult to propose $d\varepsilon _t /dq$.
Negative (positive) value denotes surface (volume) resonance.

\begin{figure}
\centerline{\includegraphics[width=11cm]{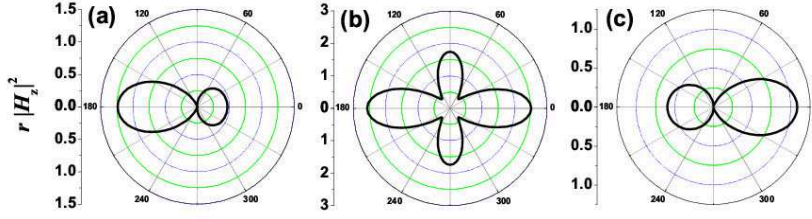}}
\vspace{-2ex}\caption{(Color online) Polar diagrams for light
intensity versus observation angle $ \theta$ associated with Fano
resonance and fast-switching radiation pattern. We assume $q = 0.1$
and $\varepsilon _r = - 1$. (a) $\delta  =  - 0.0065$; (b) $\delta =
- 0.0067835$; (c) $\delta  = - 0.0071$. The switching radiation
pattern versus the change of birefringence has been demonstrated
(Media 1).} \label{fig:2}
\end{figure}

Strong variations of the total scattered intensity are caused by a
tiny change in birefringence. As is shown in Fig.~\ref{fig:2}, when
the parameters $\varepsilon _r $ and $\varepsilon _t $ fit the
condition of the surface plasmon resonance, an equilibrium state
with a symmetrical pattern is yielded, and in the vicinity of the
equilibrium, the radiation pattern flips fast just because of a tiny
perturbation of the birefringence ( here we change $\varepsilon _t $
while keeping $ \varepsilon _r = -1$). With the increase in
$\varepsilon _t  $, more and more light is directed into the forward
direction ``before'' it passes the equilibrium state; but after
right passing that symmetrical state, light starts to be directed
back into the backward direction, and such switch is quite sensitive
to the variation of birefringence.

\begin{figure}
\centerline{\includegraphics[width=9cm]{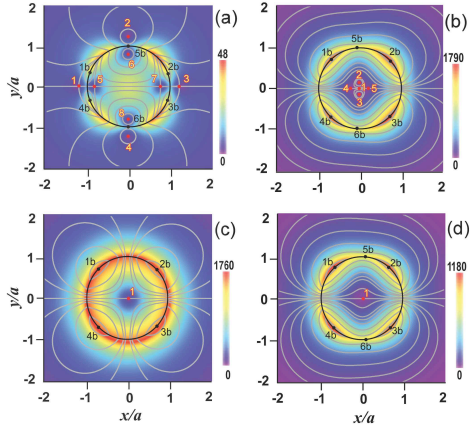}}
\vspace{-2ex}\caption{(Color online) Poynting vector $|{\bf S}|$
distribution, lines of Poynting vector ${\bf S}$, and singular
points for (a) isotropic cylinder $\delta  = 0$ and anisotropic ones
(b) $\delta   =  - 0.0065$; (c) $\delta  = - 0.0067835$; (d) $\delta
= - 0.0071$. Parameters: $q = 0.1$ and $\varepsilon _r = - 1$. In
the figures two types of singularities are shown: usual
singularities (red points) and boundary singularities (black points
at the cylinder boundary). The sensitive dependence of singularities
along the change of anisotropy can be seen (\textbf{Media 2}).}
\label{fig:3}
\end{figure}

To reveal the near-field distribution when Fano resonance occurs, we
investigate Poynting vector ${\bf S}$ bifurcation and distribution
of singularity points which is sensitively altered by the
birefringence of the particle. In Fig. \ref{fig:3}, we investigate 4
typical situations: 1) no birefringence is present (Fig.~3(a)); 2)
birefringent case at equilibrium resonant state (in Fig.~3(c)); 3)
birefringent cases ``before'' and ``after'' the equilibrium resonant
state (in Figs.~3(b,d) respectively). Based on our approach and
analysis, the Poynting vector lines and their bifurcations
demonstrate various types of singular points: (1) ordinary
singularities defined as the zeroes of the vector ${\bf S}$; (2)
so-called boundary singular points defined as $S_r(r=a,\varphi)=0$.
It shows that the ordinary singularities (red dots in \textbf{Media
2}) outside the particle will be shifted further away (out of the
plot area) but those inside the particle will be moving towards the
center and gradually merging into one common singularity in the
center eventually.

The number of boundary singularities keeps unaffected when the
object becomes anisotropic, except at the equilibrium resonant
state. However, the positions of those boundary singularities will
be sensitively dependent on the degree of anisotropy. The points
1b--4b appear for all cases, where $b$ denotes boundary singular
points. Points 5b and 6b are more interesting. If no anisotropy
exists as in Fig.~3(a), 5b and 6b are located on the crossing points
of the object's boundary and the vertical line passing through the
center (0,0). When radial anisotropy is introduced as in
Figs.~3(b,d), points 5b and 6b will be shifted away from the
centered vertical line. Of particular interest is the situation of
Fig.~3(c), where a symmetrical radiation pattern holds in both near
and far regions. In such an equilibrium state, points 1b-4b are
located symmetrically on the boundary while Points 5b and 6b
disappear. It is found that the change of singular points is not
continuous in birefringent situations, and points 5b and 6b are
either shifted away from the center line or not present (see the
black dots in \textbf{Media 2}).

Figure~3(a) also reveals that there are eight normal singularities
in the vicinity of the isotropic cylinder (more points exist afar).
The points 1--4 (5--8) are situated outside (inside) the cylinder.
Birefringent cylinder offers the fundamental distinction: a saddle
point arises exactly at the center (0,0). The numbers of singular
points become less near a birefringent cylinder, but some points
exist outside the ranges of the figures. The singularity
investigation gives us more physical insights of how energy is
directed and localized, which may be helpful in exploring
calculation of heating, radiation pressure and trapping furthermore.

\begin{figure}
\centerline{\includegraphics[width=9.2cm]{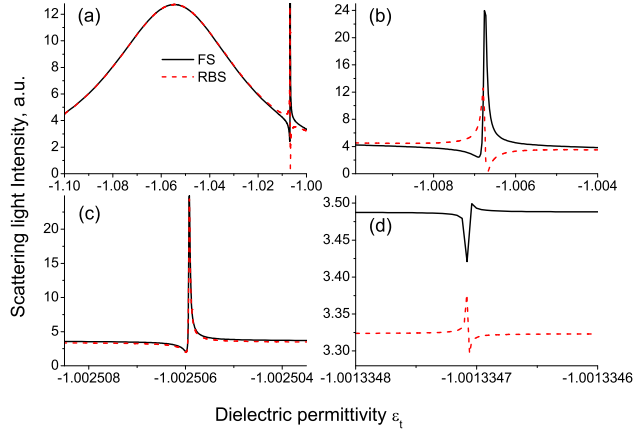}}
\vspace{-4ex}\caption{(Color online) The far-field scattering
intensity (a.u.) versus transversal permittivity $\varepsilon _t$
when $q = 0.1$ and $\varepsilon _r = - 1$. (a) Forward scattering
(FS) and radar backscattering (RBS) in the region of
$-1.1<\varepsilon _t<-1$, i.e., $-0.1<\delta<0$. (b) Asymmetric
resonance corresponding to fast-switching radiation pattern near
$\varepsilon _t\approx-1.0068$ shown in Fig.~3. (c) Symmetric
resonance near $\varepsilon _t\approx-1.002506$ which leads to
almost symmetrical radiation pattern (see the evolution in
\textbf{Media 3}). (d) Symmetric resonance near $\varepsilon
_t\approx-1.0013347$ which is less dominant. } \label{fig:4}
\end{figure}

To have a better understanding of the birefringence-induced Fano
resonance, we investigate birefringence dependence of the far-field
scattering intensity, particularly for forward scattering (FS) and
radar backscattering (RBS). In Fig.~4(a), it is clear that there are
very sensitive resonances in the narrow region of $-1.01<\varepsilon
_t<-1$. Specifically, we found three resonances as shown in
Fig.~4(b), 4(c) and 4(d), respectively. The fast-switching radiation
(i.e., FS and RBS flip over), which attributes to similar Fano
resonance in Figs.~2 and 3, is corresponding to the first resonance
reported in Fig.~4(b). It can be found in \textbf{Media 1} that such
resonance is quite asymmetric. On the contrary, \textbf{Media 3}
clearly demonstrates that the resonance in Fig.~4(c) is quite
symmetric along the change of birefringence near
$delta\approx-0.002506$. The additional resonance in Fig.~4(d)
appears to be also asymmetric but less dominant compared to the
other resonances, which is not discussed herein.

Does it mean that the last sentence above is not true after your
recent checks (your previous comments were as below)? If it is still
true, i think the current Fig.5 is not 100\% targeted to the
reviewer's concern. Fig.5 is great, but presents too much
information (without matching discussion) and didn't selectively
address reviewer's question.

We only need to show two cases (e.g., $\alpha=0.01$ and
$\alpha=0.1$) or just only one case $\alpha=0.01$, if we can exactly
illustrate the process of first resonance being broadened and one
high-order resonance being narrower, and we can tune different
$\epsilon_r$ to let it pronounce.

The birefringence-induced Fano resonance with extreme sensitivity is
presented and can be used for different applications, e.g., data
storage and optical recording. For more detailed information, we
investigate the both near-field and far field phenomena by using the
full-wave approach. The localized field distributions in the
vicinity of the resonant condition agrees with the directivity
switch in far-field due to the Fano resonance and high-order modes'
interference. Such Fano resonance is found to be existing even when
the loss is not negligible or the particle's size is large. In
conclusion, Fano resonances are generated with the condition
$\varepsilon _t< 1/\varepsilon _r $, which leads to a giant switch
in the principal light pole induced near surface plasmon resonance
with a tiny perturbation in birefringence.

Media 1, 2, and 3 are movies, supplied as supplementary files. This
work was supported by the National University of Singapore under
Grant No. R-263-000-574-133.


\begin{thebibliography}{99}

\bibitem{Tsai} W.L. Barnes, A. Dereux, and T.W. Ebbesen, Nature \textbf{424}, 824
(2003). S.A. Maier \emph{et al.}, Nat. Mater. \textbf{2}, 229
(2003).
\bibitem{Bohren} C. F. Bohren and D. R. Huffman, Absorption and Scattering of
Light by Small Particles (Willey, New York, 1983). H. C. Van De
Hulst, Light Scattering by Small Particles (Dover, New York, 2000.)




\bibitem{Tribelsky} M. I. Tribelsky and B. S. Luk'yanchuk, Phys. Rev. Lett. \textbf{97}, 263902 (2006).

\bibitem{Luk'yanchuk1} B. S. Luk'yanchuk and V. Ternovsky, Phys. Rev. B \textbf{73}, 235432
(2006).




\bibitem{Qiu} C.-W. Qiu and B.S. Luk'yanchuk, J. Opt. Soc. Am. A \textbf{25}, 1623 (2008).

\bibitem{Maier_ACS09} F. Hao \emph{et al.}, ACS Nano \textbf{3}, 643 (2009).

\bibitem{Giessen09} N. Liu \emph{et al.}, Nat. Mater. \textbf{8}, 758 (2009).

\bibitem{Tribelsky2} M. I. Tribelsky \emph{et al.}, Phys. Rev. Lett. \textbf{100}, 043903 (2008).

\bibitem{Boris_NM} B. Luk`yanchuk \emph{et al.}, Nat. Mater. \textbf{9}, 707 (2010).



\bibitem{Maier_NL} N. Verellen \emph{et al.}, Nano Lett. \textbf{9}, 1663 (2009).

\bibitem{Stockman} M.I. Stockman, S.V. Faleev, and D.J. Bergman, Phys. Rev. Lett. \textbf{87}, 167401 (2001).

\bibitem{Garcia} F. J. Garcia-Vidal, L. Martin-Moreno, T. W. Ebbesen,
and L. Kuipers, Rev. Mod. Phys. \textbf{82}, 729 (2010).

\bibitem{Pendry} J.B. Pendry, D. Schurig, and D.R. Smith, Science \textbf{312}, 1780 (2006).


\bibitem{Lucas} A.A. Lucas, L. Henrard, Ph. Lambin, Phys. Rev. B \textbf{49}, 2888 (1994).

\bibitem{Sten} J.C.E. Sten, IEEE Trans. Dielectr. Electr. Insul. \textbf{ 2}, 360 (1995).

\bibitem{Lange} B. Lange and S.R. Aragon, J. Chem. Phys. \textbf{92}, 4643 (1990).

\bibitem{Peterlin} P. Peterlin, S. Svetina, and B. Zeks, J. Phys.: Condens. Matter \textbf{19}, 136220 (2007).

\bibitem{Sukhorukov} V.L. Sukhorukov, G. Meedt, M. Kurschner, U. Zimmermann, J. Electrost. \textbf{50}, 191 (2001).

\bibitem{Ambjornsson}  T. Ambjornsson, G. Mukhopadhyay, S.P. Apell, and M. Kall,  Phys. Rev. B \textbf{73}, 085412 (2006).


\bibitem{Fung} L. Gao \emph{et al.}, Phys. Rev. E \textbf{78}, 046609
(2008). Y. Wu \emph{et al.}, Phys. Rev. B \textbf{74}, 085111
(2006).


\end{thebibliography}
\end{document}